\documentclass[article,nojss,shortnames]{jss}

\usepackage[]{graphicx}
\usepackage[]{color}
\makeatletter
\def\maxwidth{ %
  \ifdim\Gin@nat@width>\linewidth
    \linewidth
  \else
    \Gin@nat@width
  \fi
}
\makeatother

\definecolor{fgcolor}{rgb}{0.345, 0.345, 0.345}

\usepackage{framed}
\makeatletter
 {\par\unskip\endMakeFramed%
 \at@end@of@kframe}
\makeatother

\definecolor{shadecolor}{rgb}{.97, .97, .97}
\definecolor{messagecolor}{rgb}{0, 0, 0}
\definecolor{warningcolor}{rgb}{1, 0, 1}
\definecolor{errorcolor}{rgb}{1, 0, 0}
\newenvironment{knitrout}{}{} 

\usepackage{alltt}

\usepackage{thumbpdf}
\usepackage{amsfonts,amstext,amsmath,amssymb,amsthm}
\usepackage{accents}
\usepackage{color}
\usepackage{rotating}
\usepackage{verbatim}
\usepackage{dcolumn}
\usepackage{booktabs}

\renewcommand{\Prob}{\mathbb{P}}

\newcommand{\rY}{Y}
\newcommand{\rX}{\mX}

\newcommand{\ry}{y}
\newcommand{\rx}{\xvec}

\newcommand{\pY}{F_\rY}

\newcommand{\h}{h}

\newcommand{\basisy}{\avec}
\newcommand{\bern}[1]{\avec_{\text{Bs},#1}}

\newcommand{\parm}{\varthetavec}
\newcommand{\eparm}{\vartheta}

\newcommand{\shiftparm}{\betavec}
\newcommand{\eshiftparm}{\beta}

\newcommand{\ie}{\textit{i.e.}~}

\newcommand{\Ex}{\mathbb{E}}
\newcommand{\RR}{\mathbb{R}}

\usepackage{dsfont}

 \DeclareMathOperator{\ND}{N}

\def \avec {\text{\boldmath$a$}}

\def \xvec {\text{\boldmath$x$}}    \def \mX {\text{\boldmath$X$}}

\def \betavec         {\text{\boldmath$\beta$}}

\def \varthetavec     {\text{\boldmath$\vartheta$}}

\author{Torsten Hothorn \\ Universit\"at Z\"urich}
\Plainauthor{Hothorn}

\title{Top-down Transformation Choice}
\Plaintitle{Top-Down Transformation Choice} 
\Shorttitle{Top-Down Transformation Choice}

\newcommand{\age}{\text{age}}
\newcommand{\BMI}{\text{BMI}}
\newcommand{\sex}{\text{sex}}
\newcommand{\smoking}{\text{smoking}}

\newcommand{\expit}{\text{expit}}
\newcommand{\oBMI}{O_{\BMI}}
\newcommand{\pBMI}{F_{\BMI}}
\newcommand{\qBMI}{Q_{\BMI}}
\newcommand{\dBMI}{f_{\BMI}}

\newcommand{\apriori}{\textit{a priori}~}

\Abstract{

Simple models are preferred over complex models, but over-simplistic models
could lead to erroneous interpretations.  The classical approach is to start
with a simple model, whose shortcomings are assessed in residual-based model
diagnostics.  Eventually, one increases the complexity of this initial
overly simple model and obtains a better-fitting model.  I illustrate how
transformation analysis can be used as an alternative approach to model
choice.  Instead of adding complexity to simple models, step-wise complexity
reduction is used to help identify simpler and better-interpretable models. 
As an example, body mass index distributions in Switzerland are modelled by
means of transformation models to understand the impact of sex, age, smoking
and other lifestyle factors on a person’s body mass index.  In this process,
I searched for a compromise between model fit and model interpretability. 
Special emphasis is given to the understanding of the connections between
transformation models of increasing complexity.  The models used in this
analysis ranged from evergreens, such as the normal linear regression model
with constant variance, to novel models with extremely flexible conditional
distribution functions, such as transformation trees and transformation
forests.

}

\Plainkeywords{Transformation analysis, conditional transformation model, 
          conditional distribution function, conditional quantile function, 
          distribution regression, stratified linear transformation model,
          body mass index}

\Address{
  Torsten Hothorn\\
  Institut f\"ur Epidemiologie, Biostatistik und Pr\"avention \\
  Universit\"at Z\"urich \\
  Hirschengraben 84, CH-8001 Z\"urich, Switzerland \\
  \texttt{Torsten.Hothorn@uzh.ch} \\

}
\IfFileExists{upquote.sty}{\usepackage{upquote}}{}

\begin{document}

\section{Introduction}

Let's face it.  The work of statisticians is considered boring in the public
eye.  Nobody publishes page turners on the thrilling aspects of data
analysis, yet the quest for a good model can be as exciting as
detective work.  One of my favourite paperback characters is LAPD detective
Harry Bosch in the crime novels of Michael Connelly. Like
Harry, who follows the traces left by the murderer on the crime scene to form a
theory about the culprit, the experienced data analyst follows the traces
left by the data-generating process in the residuals of an over-simplistic
model.  Unlike Harry, who of course always succeeds in arresting the
murderer, the statistician can never be sure whether the correct or even an approximately useful model was found.  In the quests for a suspect or for
a good model, parsimonious explanations are preferred by Occam's razor. 
Therefore, in residual-based model diagnostics, the data analyst starts with
a very simple model, whose complexity is increased by step-wise refinement
until all signs of lack of fit disappear from the residuals.  I refer to
such a procedure as ``bottom-up model choice'' because one moves from simple
to more complex models.  In this tutorial, I consider moving in the
opposite direction, \ie from complex to simple models, for distributional
regression.  This ``top-down approach'' to model choice begins with the most 
complex model that one can come up with that explains both signal and noise without overfitting the
data. In a regression setup, such a model would describe as accurately as possible
the conditional distribution of the response given the explanatory variables.
Once such a model is established as a benchmark for comparison with simpler
models, one can start to reduce model complexity step-wise. 
In the crime novel scenario, the top-down data analyst takes the role of an
eyewitness at the scene.  What one ``sees'' in this process is, of
course, still a portrayal and not the real thing.  There is no way to
``see'' the correct model.  In top-down model choice, however, the
trajectories through model space will be guided by assessments of vital
models.  In bottom-up model choice, by contrast, the horizon is limited by the
amount of information that one can find in traces in deceased models.

In this tutorial, I focus on top-down model choice in continuous regression
problems.  Conceptually, a regression model is a family of conditional
distributions for some response $\rY$ given a specific configuration of
explanatory variables $\rX = \rx$.  The model describes both signal and
noise, \ie the variability explained by the explanatory variables and the
unexplained variability.  Unfortunately, this point of view only applies to
relatively simple models that assume a certain parametric distribution, whose
parameters partially depend on the explanatory variables.  So-called
``non-parametric regression models'' \citep{Fahrmeir_Kneib_Lang_2013} 
often restrict their attention to the
signal $\Ex(\rY \mid \rX = \rx) = m(\rx)$, with non-linear
conditional mean function $m$, while treating the noise, \ie all
higher moments of the conditional distribution, as a nuisance or essentially ignoring it. 
Such procedures, for example random forests \citep{Breiman_2001}, are
extremely powerful when estimating complex conditional mean functions.
However, one cannot infer the entire conditional distribution using random
forests or similar methods.  This renders top-down model
choice impossible because reductions in complexity require switching
between different model classes or even crossing the borders between the
parametric and non-parametric empires.  The comparison of two models from
different classes is difficult, and thus it is difficult to decide whether the simpler
model is more appropriate than the more complex one.

The implementation of top-down model choice is much simpler when the most
complex and the most simple model are members of the same family. 
Conditional transformation models from the transformation family of
distributions
\citep{Hothorn_Kneib_Buehlmann_2014,Hothorn_Moest_Buehlmann_2016} include
many important established off-the-shelf regression models.  In addition,
tailored models can be created, \textit{in vivo} with our brains 
and \textit{in silico} using open-source software, which allow smooth transitions between models of
different complexity.  In a nutshell, the class of conditional
transformation models
\begin{eqnarray*}
\Prob(\rY \le \ry \mid \rX = \rx) = \pY(\ry \mid \rx) = F(\h(\ry \mid \rx))
\end{eqnarray*} 
assumes that the conditional distribution function $\pY$ of $\rY$ given $\rX
= \rx$ can be written as the composition of an \apriori specified continuous
cumulative distribution function $F$ and some conditional transformation
function $\h(\ry \mid \rx)$.  The latter function monotonically increases in its first argument for each $\rx$. It is important to note that the entire
conditional distribution, and not just its mean, is modelled by $\h$. In
this sense, and unlike common regression models, there is no decomposition into 
signal (the conditional mean) and noise (the remaining higher moments) in this class of
transformation models. Changing model complexity means
changing the complexity of the conditional transformation function $\h$, and
I thus refer to top-down model choice in conditional transformation models
as top-down transformation choice.

Model complexity in the class of conditional transformation models is linked
to smooths $\h(\ry \mid \rx)$ of varying complexity with respect to $\ry$. 
These conditional transformation functions may vary with the explanatory
variables $\rx$ in arbitrary ways, including interactions and
non-linearities.  In this paper, I consider model choice as an art rather
than an exact science.  No formal algorithm leading to an ``optimal'' model
will be presented.  Instead, I argue that the possibility of modelling a
cascade of decreasingly complex conditional distribution functions in the
same model class gives us new possibilities to investigate goodness of fit
or lack thereof.  A fair amount of subjectivity will remain in this process,
as is always the case in classical model diagnostics.  I shall be less
concerned with the technical subtleness of parameter estimation in the
models discussed here and refer the reader to more formal results published
elsewhere.  Instead, I illustrate practical aspects of top-down
transformation choice by a tour-de-force through transformation models
describing the impact of lifestyle parameters, such as smoking or physical
activity, on the body mass index (BMI) distribution in the Swiss population.

I will proceed by introducing the Swiss Health Survey and the variables 
dealt with in Section~\ref{section:SHS}.  In a very simple setup, I first
illustrate a bottom-up route, starting with a normal linear model and ending
with a more complex non-normal transformation model, for describing the BMI
distribution of females and males at various levels of smoking
(Section~\ref{section:sexsmoking}).  I then try to reduce the complexity
again until an interpretable model that fits the data roughly as well as the most complex model can be found.  In addition to a consideration of sex and smoking, I consider
age and some lifestyle variables in a more realistic setup of top-down
transformation choice in Section~\ref{section:ctm}.

\section{Body Mass Index in the Swiss Health Survey} \label{section:SHS}

The Swiss Health Survey (SHS) is a population-based cross-sectional survey. 
It has been conducted every five years since 1992 by the Swiss Federal
Statistical Office \citep{SHS_2012}.  
For this tutorial, I restricted the sample to $16{,}427$ individuals aged
between $18$ and $74$ years from the 2012 survey.  
Study samples were obtained by stratified random
sampling using a database with all private household landline telephone
numbers.  Data were collected by telephone interviews and self-administered
questionnaires.  
Height and weight were self-reported in telephone interviews.  Observations 
with extreme values of height and weight were excluded (highest and lowest
percentile by sex).  Smoking status was categorised into never, former,
light ($1-9$ cigarettes per day), moderate ($10-19$) and heavy smokers
($>19$).  Never smokers stated that they did not currently smoke and never
regularly smoked longer than six months; former smokers had quit smoking 
but have smoked for more than six months during their life course.  One cigarillo 
or pipe counted as two cigarettes, and one cigar counted as four cigarettes.  The 
following lifestyle variables were included 
and assessed by telephone interview and self-administered questionnaire: 
fruit and vegetable consumption, physical 
activity, alcohol intake, level of education, nationality and place of
residence. Fruit and
vegetable consumption was combined in one binary variable that comprised the
information on whether both fruits and vegetables were consumed daily or
not.  The variable describing physical activity was defined as the number of
days per week a subject started to sweat during leisure time physical
activity and was categorised as $>2$ days, $1-2$ days and none.  Alcohol
intake was included using the continuous variable gram per day.  Education
was included as highest degree obtained and was categorised into mandatory
(International Standard Classification of Education, ISCED 1-2), secondary
(ISCED 3-4), and tertiary (ISCED 5-8) \citep{UNESCO_2012}.  Nationality
had the two categories Swiss and foreign.  Language reflected
cultural and regional differences within Switzerland, and the three categories
German/Romansh, French and Italian were taken into account. Sampling weights
of this representative survey were considered for the estimation of
all models reported in this tutorial. More detailed information about
this study and an analysis using simple transformation models is given in
\cite{Lohse_Rohrmann_Faeh_2017}.

\section{Sex- and Smoking-specific BMI Distributions}
\label{section:sexsmoking}

\begin{figure}[t]
\begin{knitrout}
\definecolor{shadecolor}{rgb}{0.969, 0.969, 0.969}\color{fgcolor}
\includegraphics[width=\maxwidth]{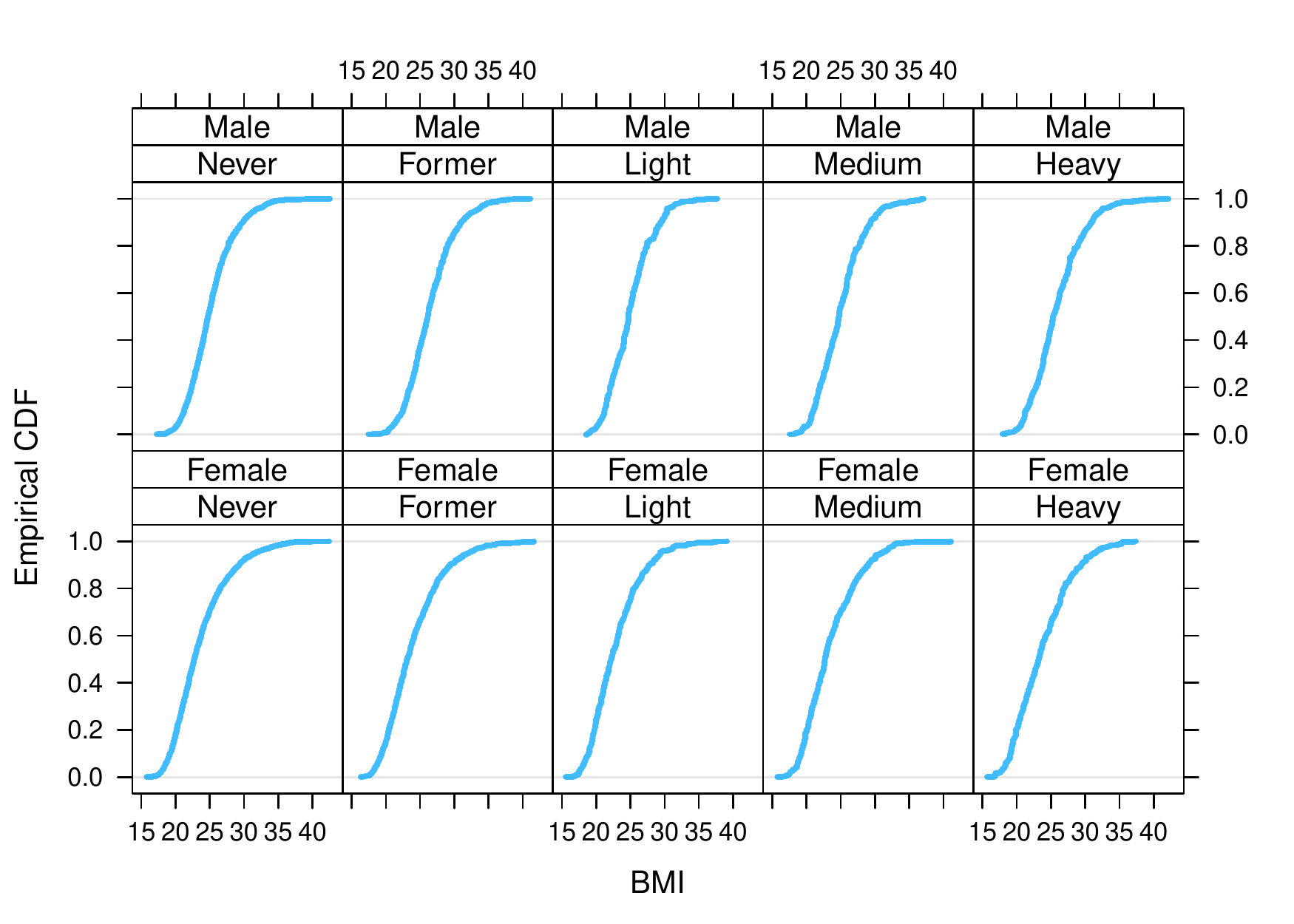} 

\end{knitrout}
\caption{The empirical cumulative distribution function (CDF) 
         of BMI given sex and smoking. For each combination of
         sex and smoking, the weighted empirical CDF taking sampling
         weights into account is presented. 
         \label{ss-ecdf}}
\end{figure}

I start with the very simple situation where the conditional
distribution of BMI depends on sex and smoking only.  Smoking was assessed on five
different levels (never smoked, former smokers, light smokers, medium
smokers and heavy smokers). Therefore, I am interested in the conditional
distribution of BMI in these $10$ groups of participants. 
Figure~\ref{ss-ecdf} presents the empirical cumulative distribution
functions, \ie the non-parametric maximum-likelihood estimators for the
underlying continuous distributions, for each of the $10$ combinations of
sex and smoking.  At the same time, the plot also represents the
uncompressed raw data.  With a high-enough resolution, one could recover the
original BMI values and the corresponding sampling weights from such an image. 
Consequently, goodness of fit can be assessed by overlaying the empirical
cumulative distribution functions with their model-based counterparts in
this simple setup.  I will try to find a suitable parametric model this
way.  In addition to this rather informal approach, I will study the
increase of the log-likelihoods as model complexity is increased.
In the classical bottom-up approach, one would start with a very
simple model assuming conditional normal distributions.  The next section
discusses possible choices in this model class.

\subsection{Normal Models}

The normal cell-means model with constant variance \begin{eqnarray}
\label{mod:ss1} \BMI \mid \sex, \smoking \sim \ND(\mu(\sex:\smoking),
\sigma^2) \end{eqnarray} assumes normal distributions with a common variance
for all conditional BMI distributions.  Means are allowed to vary between
the groups defined by sex and smoking.  The notation $\mu(\sex:\smoking)$
indicates that the conditional mean is specific to each combination of sex
and smoking in this cell-means model, \ie there are a total of $10$ parameters
$\mu(\sex:\smoking)$. With a residual standard error of
$\hat{\sigma} = 3.73$, a log-likelihood of
$-47010.46$ was obtained, and the estimated cell-means, with $95\%$ confidence
intervals, are shown in Table~\ref{ss1-tab}.

\begin{table}[t]
\begin{center}
\caption{Normal cell-means model (\ref{mod:ss1}) with constant variance. 
         Estimated means of BMI for each
         combination of sex and smoking, with $95\%$ confidence 
         intervals are shown. \label{ss1-tab}}
\begin{tabular}{llrr}
\toprule
 && \multicolumn{2}{c}{Sex}\\
\cmidrule{3-3}\cmidrule{4-4}
Smoking && \multicolumn{1}{c}{Female}&\multicolumn{1}{c}{Male}\\
\midrule
Never  && 23.57 (23.47--23.68) & 25.18 (25.06--25.30)\\
Former && 23.94 (23.76--24.12) & 26.46 (26.30--26.62)\\
Light  && 22.85 (22.59--23.11) & 25.00 (24.73--25.27)\\
Medium && 23.45 (23.18--23.72) & 25.00 (24.75--25.26)\\
Heavy  && 23.72 (23.36--24.08) & 25.90 (25.65--26.15)\\
\bottomrule
\end{tabular}

\end{center}
\end{table}

\begin{figure}[t]
\begin{knitrout}
\definecolor{shadecolor}{rgb}{0.969, 0.969, 0.969}\color{fgcolor}
\includegraphics[width=\maxwidth]{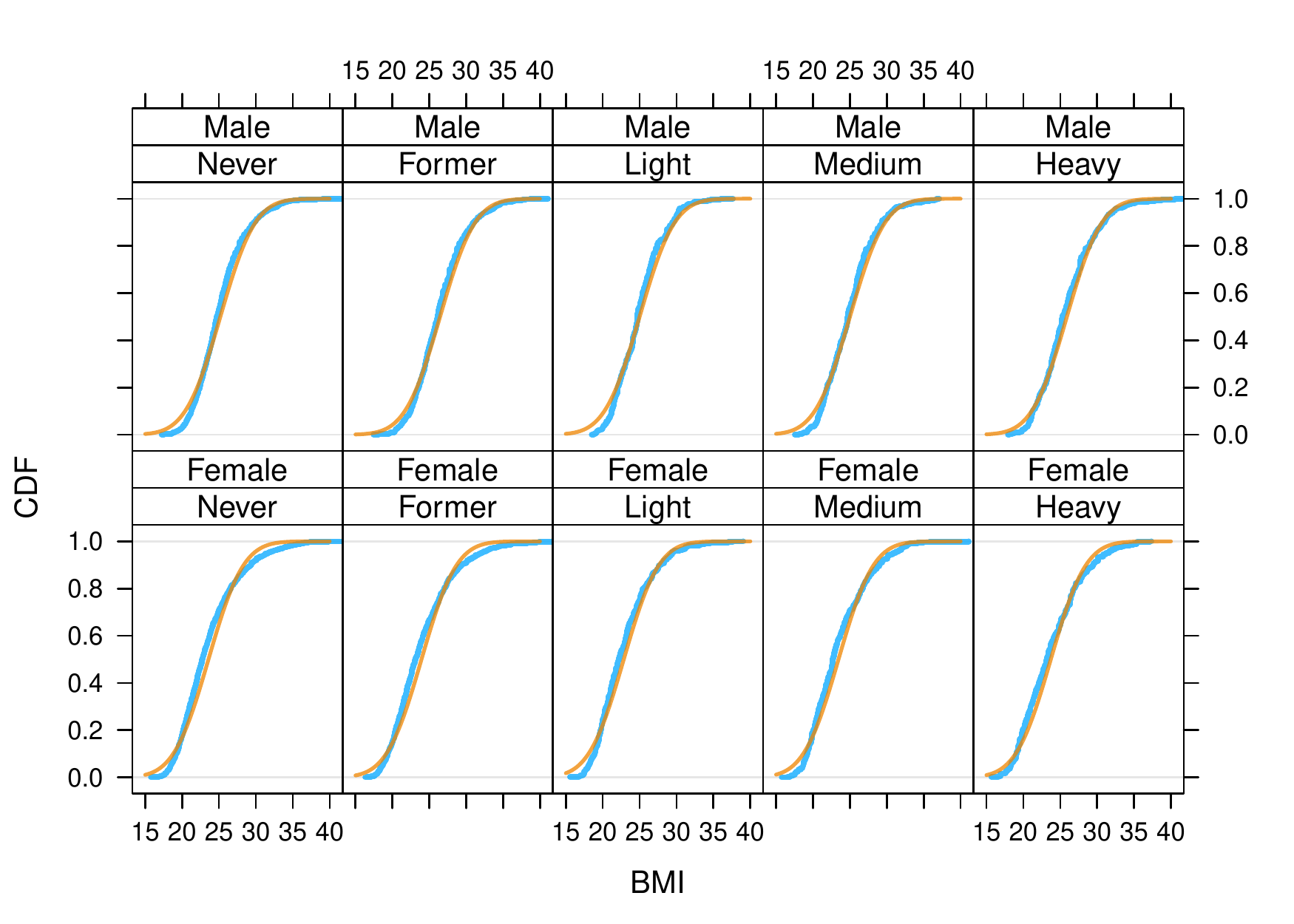} 

\end{knitrout}
\caption{Normal cell-means model (\ref{mod:ss1}) with constant variance. The 
         empirical (blue) and model-based (yellow) cumulative
         distribution functions (CDF) of BMI given sex and smoking
         are shown. \label{ss1-plot}}
\end{figure}

How well does this model fit the data? I want to answer this question by graphically
comparing the conditional distribution functions obtained from this model to
the corresponding empirical conditional distributions, and thus the raw
data. The model-based conditional cumulative distribution functions
\begin{eqnarray*} 
\Prob(\BMI \le \ry \mid \sex, \smoking) = \pBMI(\ry \mid \sex, \smoking) = \Phi\left(\frac{\ry - \mu(\sex:\smoking)}{\sigma}\right)
\end{eqnarray*}
overlay the empirical cumulative distribution functions in Figure~\ref{ss1-plot}. 
While not being completely out of line, the considerable differences between the empirical
and model-based distribution functions certainly leave room for improvement.
An obvious increase in the complexity allows for group-specific
variances in the model
\begin{eqnarray} \label{mod:ss2}
\BMI \mid  \sex, \smoking \sim \ND(\mu(\sex:\smoking), \sigma(\sex:\smoking)^2).
\end{eqnarray}

\begin{figure}[t]
\begin{knitrout}
\definecolor{shadecolor}{rgb}{0.969, 0.969, 0.969}\color{fgcolor}
\includegraphics[width=\maxwidth]{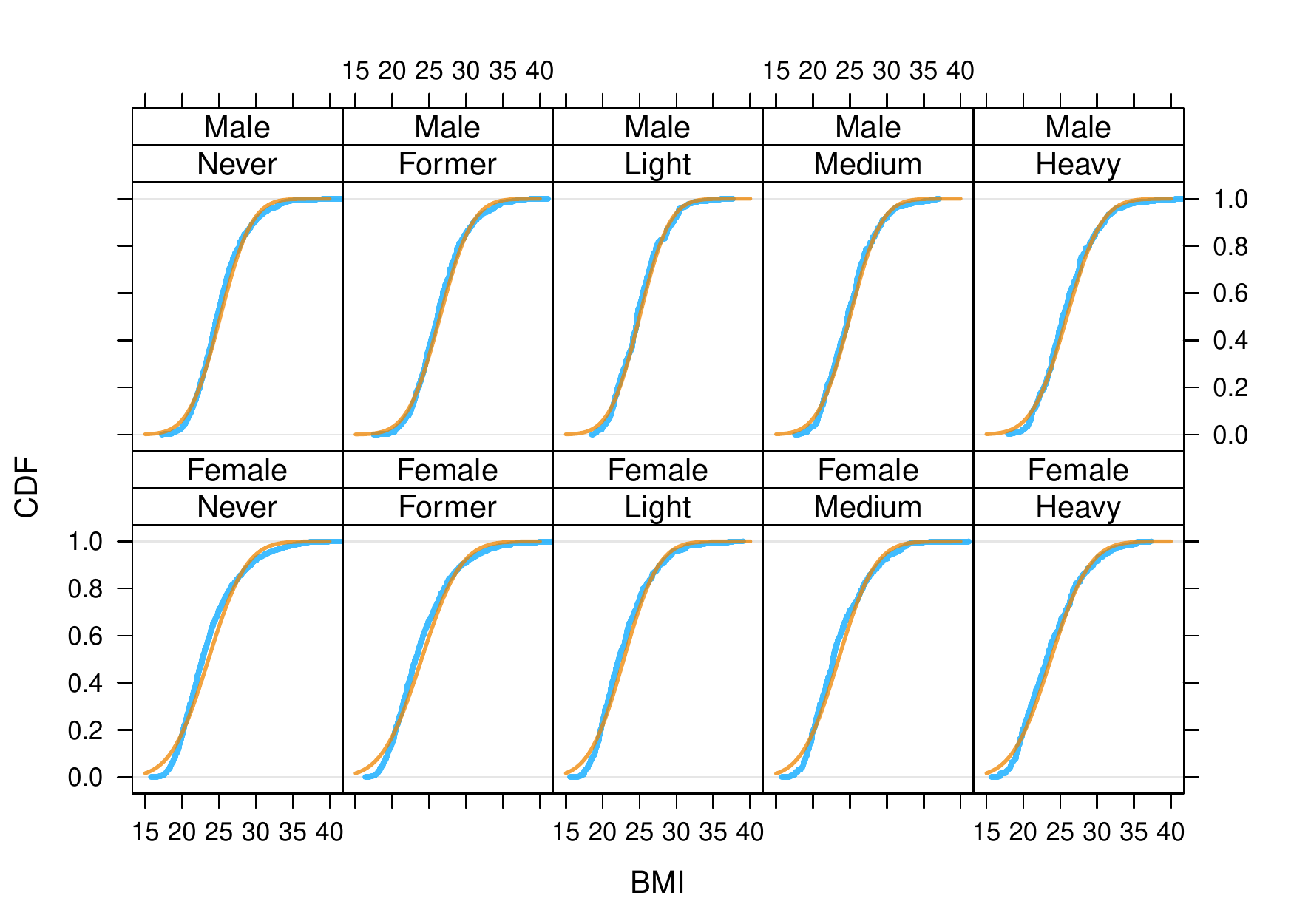} 

\end{knitrout}
\caption{Normal cell-means model (\ref{mod:ss2}) with heterogeneuous variance. The 
         empirical (blue) and model-based (yellow) cumulative
         distribution functions (CDF) of BMI given sex and smoking
         are shown. \label{ss2-plot}}
\end{figure}

The log-likelihood in this $20$-parameter model increased to
$-44801.19$, and the corresponding conditional distribution
functions in Figure~\ref{ss2-plot} were closer to the empirical cumulative
distribution functions.  For males, the model-based normal distributions
were very close to the empirical conditional BMI distributions.  For females,
however, there still was a considerable discrepancy between model and data,
especially in the lower tails.  The BMI distributions of females deviated
from normality much more than the BMI distributions of males (note that I am not
saying that males are normal and females are not!).  It is clear that one has to
move to a non-normal error model, at least for females, and the
transformation models discussed below are a convenient way to do so.

The normal models are a special case of transformation models and thus the
latter class is a very natural extension of the former. 
To see the connection, consider the conditional distribution function
\begin{eqnarray*}
\Prob(\BMI \le \ry \mid \sex, \smoking) & = & \Phi\left(\frac{\ry - \mu(\sex:\smoking)}{\sigma(\sex:\smoking)}\right)
\\
& = & \Phi\left(\eparm(\sex:\smoking) \ry - \shiftparm(\sex:\smoking)\right)
\\
& = & \Phi\left(\h(\ry \mid \sex:\smoking)\right),
\end{eqnarray*}
where $\h$ is a linear function of $\ry$ with parameters
$\eparm(\sex:\smoking) = \sigma(\sex:\smoking)^{-1}$ and
$\shiftparm(\sex:\smoking) = \mu(\sex:\smoking) / \sigma(\sex:\smoking)$. 
In more complex models, we will use the parameter $\eparm$ (or parameter 
vector $\parm$ for basis-transformed response values $\basisy(\ry)$) 
for modelling transformations $\h$ of the response $\ry$. Shift parameters
describing effects of explanatory variables only, \ie no
response-varying effects, will be denoted by $\beta$ 
or $\shiftparm$ later on. The above re-parameterisation shows, 
quite unsurprisingly, that a normal model
features a linear transformation function $\h$.  Consequently, non-normal conditional
distributions can be obtained by allowing a non-linear transformation
function for each combination of sex and smoking in the transformation 
models presented in the next section.

\subsection{Non-normal Transformation Models}

\begin{figure}[t]
\begin{knitrout}
\definecolor{shadecolor}{rgb}{0.969, 0.969, 0.969}\color{fgcolor}
\includegraphics[width=\maxwidth]{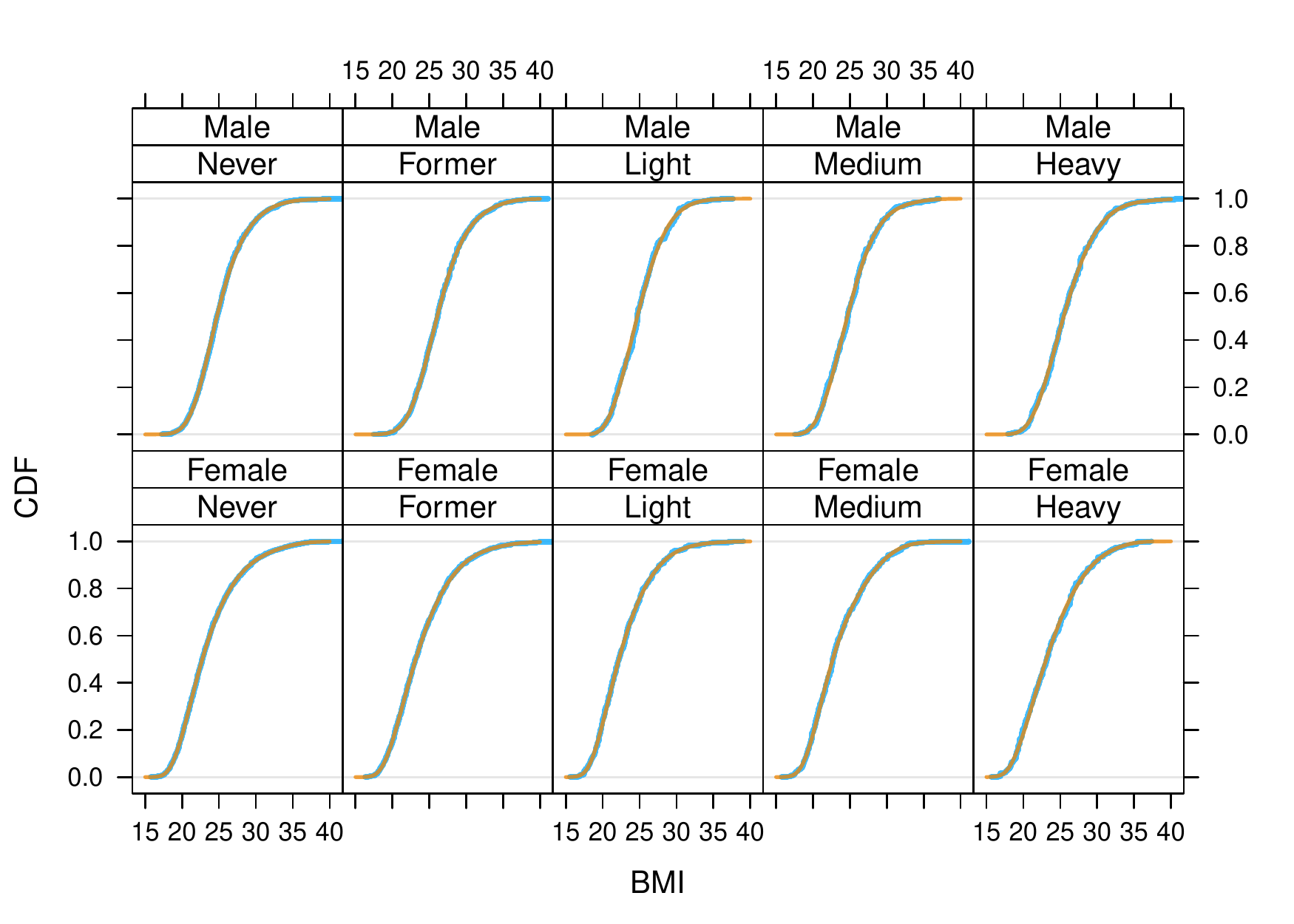} 

\end{knitrout}
\caption{Transformation model (\ref{mod:ss3}) stratified by sex and smoking. The 
         empirical (blue) and model-based (yellow) cumulative
         distribution functions (CDF) of BMI given sex and smoking
         are shown.\label{ss3-plot}}
\end{figure}

\begin{figure}[t]
\begin{knitrout}
\definecolor{shadecolor}{rgb}{0.969, 0.969, 0.969}\color{fgcolor}
\includegraphics[width=\maxwidth]{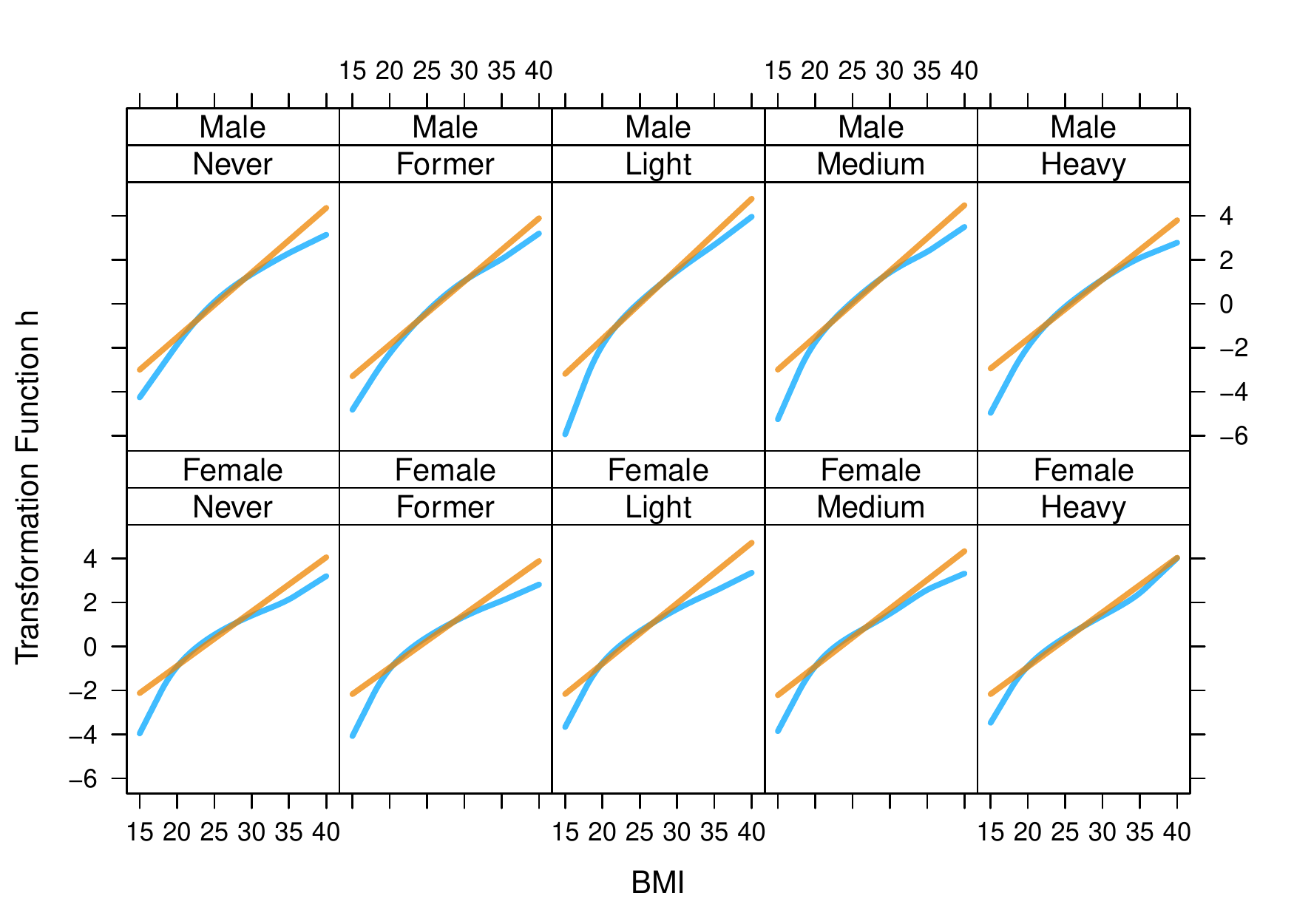} 

\end{knitrout}
\caption{Transformation model (\ref{mod:ss3}) stratified by sex and smoking. 
         Deviations from normality indicated by the non-linear transformation functions
         (blue) compared to the linear transformation functions (yellow) obtained
         from the normal cell-means model (\ref{mod:ss2}) with heterogeneous variances are shown. 
         \label{ss3-trafo-plot}}
\end{figure}

The core concept of a transformation model is a potentially non-linear
monotonically increasing transformation function $\h(\ry \mid \sex:\smoking)$,
here for each combination of sex and smoking.  For computational
convenience, I parameterised the transformation functions $\h$ in terms of 
Bernstein polynomials \citep{Farouki_2012}. 
For each of the $10$ groups, I modelled the transformation function
$\h$ by a Bernstein polynomial $\bern{5}(\ry)^\top \parm$ of
order $5$, where $\bern{5}(\ry) \in \RR^{6}$ 
are the corresponding basis functions of BMI.  A monotonically increasing Bernstein polynomial of order $5$
features $6$ monotonically increasing parameters $\parm$.  
Maximum-likelihood estimation was performed
\citep{Hothorn_Moest_Buehlmann_2016} using the \pkg{mlt} 
\citep{pkg:mlt,vign:mlt.docreg} add-on package to the 
\proglang{R} system for statistical computing \citep{R}.
With the corresponding
$60$ total parameters, the maximum log-likelihood of the model
\begin{eqnarray} \label{mod:ss3}
\Prob(\BMI \le \ry \mid \sex, \smoking) = \Phi\left(\bern{5}(\ry)^\top \parm(\sex:\smoking)\right)
\end{eqnarray}
was $-43564.30$; the notation $\parm(\sex:\smoking)$ indicates that
the parameters were estimated for each combination of sex and smoking.
One can hardly differentiate the resulting model-based conditional distribution functions 
from the empirical cumulative BMI distribution functions in
Figure~\ref{ss3-plot}.  Because a separate transformation function was
estimated for each combination of sex and smoking, this
model can be referred to as a transformation model stratified by sex and smoking.  Based on
this model, one can understand non-normality as deviation of the
transformation function from a linear function.  Figure~\ref{ss3-trafo-plot}
shows the sex- and smoking-specific transformation functions of
model (\ref{mod:ss3}) along with the
linear transformation functions obtained from the normal cell-means model
(\ref{mod:ss2}) with
heterogeneous variances.  In the centre of the distributions, the two curves
overlap, but the tails are not described well by the normal distribution. 
The differences between the two curves are more pronounced for females,
corresponding to the larger deviations from normality observed earlier.

\begin{figure}[t]
\begin{knitrout}
\definecolor{shadecolor}{rgb}{0.969, 0.969, 0.969}\color{fgcolor}
\includegraphics[width=\maxwidth]{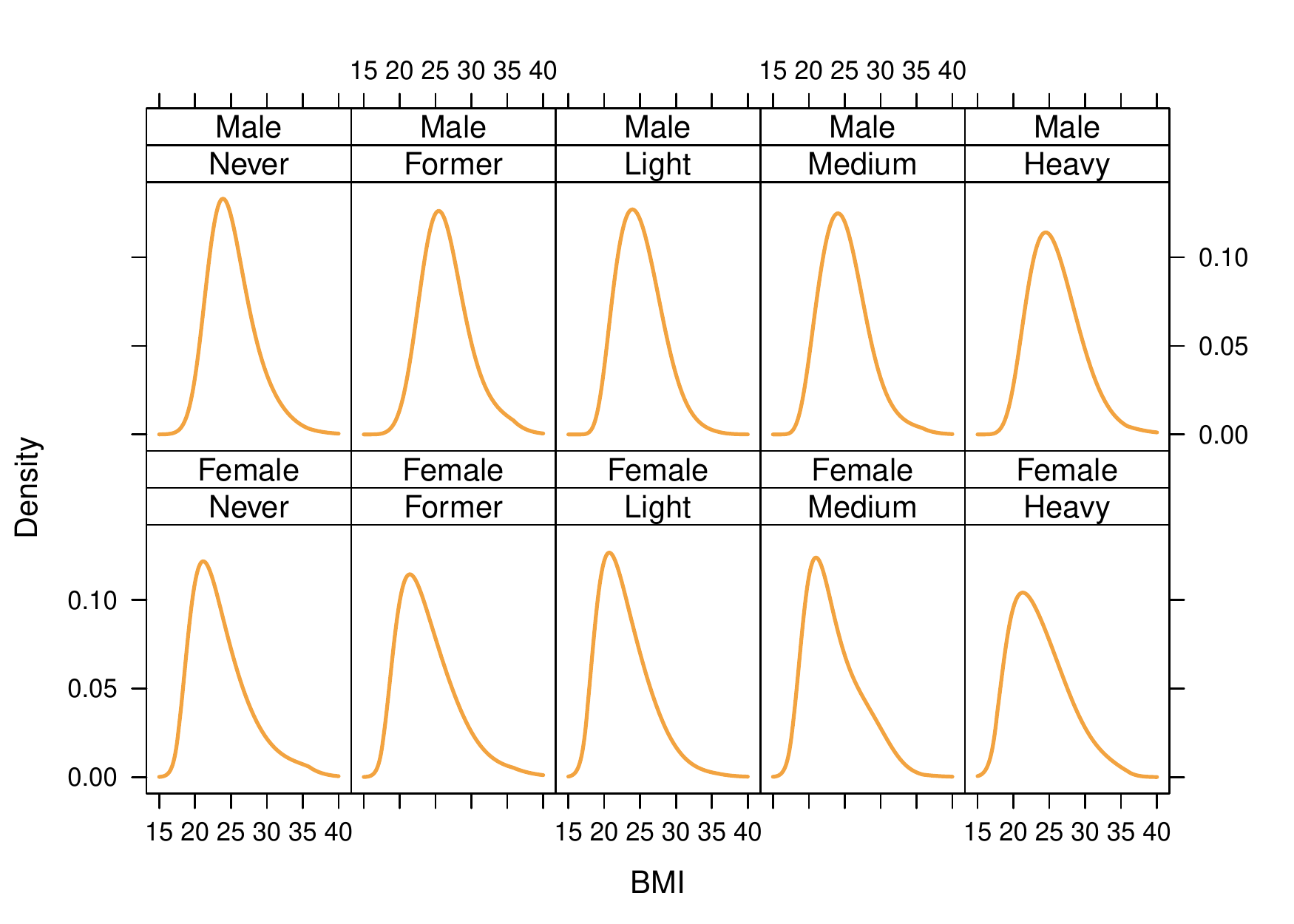} 

\end{knitrout}
\caption{Transformation model (\ref{mod:ss3}) stratified by sex and smoking. The 
         model-based conditional densities of BMI given sex and smoking
         are shown.\label{ss3-plot-density}}
\end{figure}

One nice feature of model (\ref{mod:ss3}) is the possibility to easily
derive characterisations of the distribution other than the distribution
function. Density, quantile, hazard, cumulative hazard or other
characterising functions can be derived from (\ref{mod:ss3}), and
Figure~\ref{ss3-plot-density} depicts the densities for males and females at 
the various levels of smoking. The right-skewness of the distribution, and
thus deviation from normality, was more pronounced for females. The BMI
distributions for females put more weight on smaller BMI values for females
than for males. Except for heavy smokers, the effects of smoking seemed
to be rather small.

The model fit of this stratified transformation model is now satisfactory,
as it essentially smoothly interpolates the empirical distribution functions
and thus the data in Figure~\ref{ss3-plot}.  This most complex model
describes the data well, but unfortunately, it is difficult to learn anything
from this model.  That is, one wants to understand the differences between the
conditional distributions in terms of simple parameters and not complex
non-linear functions.  A simpler model is needed.  A top-down approach to
transformation choice might help to identify a model with simpler
and interpretable transformation functions, but any necessary compromises to the model fit should not be too demanding.

Because the BMI distributions differed most between males
and females, I first simplify the model by conditioning on smoking and
stratifying by sex, \ie I introduce sex-specific transformations
$\h(\ry \mid \sex)$ and sex-specific smoking effects $\shiftparm$, the
latter being constant for all arguments $\ry$ of the conditional
distribution function, in the model
\begin{eqnarray} \label{mod:ss4}
\Prob(\BMI \le \ry \mid \sex, \smoking) = \Phi\left(\h(\ry \mid \sex) - \shiftparm(\sex:\smoking)\right).
\end{eqnarray}
This model features two transformation functions $\h(\ry \mid \text{male})$
and $\h(\ry \mid \text{female})$. For never smokers (the reference
category), these two transformation functions describe the conditional BMI
distributions, \ie
\begin{eqnarray*}
\Prob(\BMI \le \ry \mid \sex, \text{never smoked}) = \Phi\left(\h(\ry \mid \sex)\right).
\end{eqnarray*}
For the remaining smoking categories, one sex-specific parameter
$\shiftparm(\sex:\smoking)$ describes how the conditional BMI distribution
of a smoker differs from the conditional BMI distribution of a person who
never smoked by a simple shift term $\shiftparm$. Because of the ``linear''
shift term, this model could be referred to as a stratified linear
transformation model. This is, as often in statistics, a misnomer, because
the transformation $\h(\ry \mid \sex)$ of the response, \ie of the BMI values $\ry$, is
non-linear.

\begin{figure}[t]
\begin{knitrout}
\definecolor{shadecolor}{rgb}{0.969, 0.969, 0.969}\color{fgcolor}
\includegraphics[width=\maxwidth]{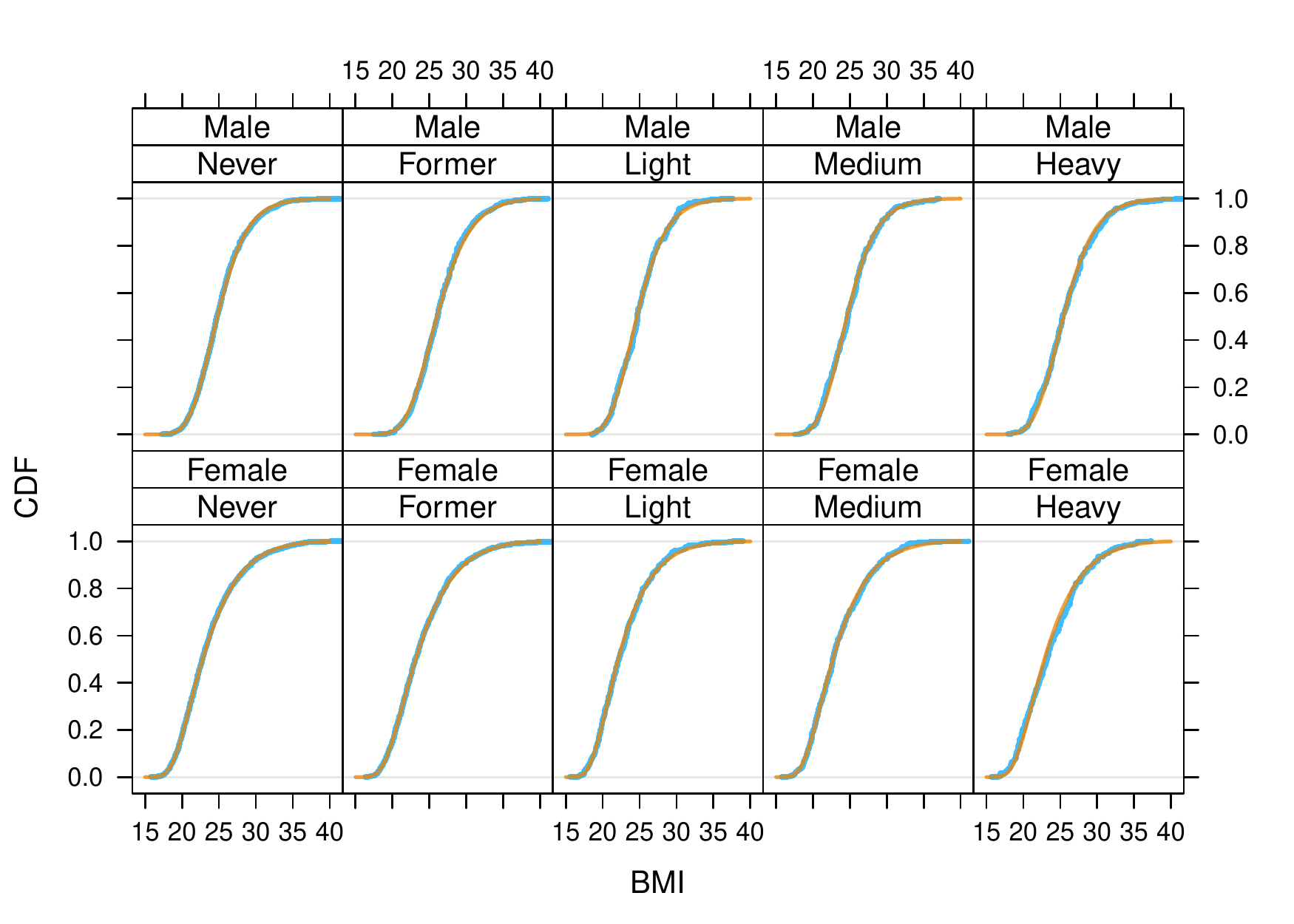} 

\end{knitrout}
\caption{Linear transformation model (\ref{mod:ss4}) with sex and smoking-specific shift, stratified by sex.
         The empirical (blue) and model-based (yellow) cumulative
         distribution functions (CDF) of BMI given sex and smoking
         are shown. \label{ss4-plot}}
\end{figure}

The log-likelihood for this model with $20$
parameters was found to be $-43602.03$, a moderate reduction
compared to the log-likelihood of the most complex transformation model
($-43564.30$).  Figure~\ref{ss4-plot} shows only minor differences
between the empirical and model-based conditional distribution functions. 
Thus, it seems that a more parsimonious model was found without
paying too high a price in terms of log-likelihood reduction.

The conceptual problem with this model, however, is lack of interpretability
of the shift term $\shiftparm(\sex:\smoking)$. In contrast to the means
$\mu$ in the normal models (\ref{mod:ss1}) and (\ref{mod:ss2}), there is
no direct interpretation of $\shiftparm$ in terms of moments of the
conditional distribution described by this model. This issue can be addressed
by changing the cumulative distribution function $F = \Phi$ of the standard
normal to the cumulative distribution function $F = \expit$ of the
standard logistic
\begin{eqnarray} \label{mod:ss5}
\Prob(\BMI \le \ry \mid \sex, \smoking) = \expit\left(\h(\ry \mid \sex) - \shiftparm(\sex:\smoking)\right).
\end{eqnarray}
When the cut-off $\ry$ is fixed, this is a logistic regression
model for the binary outcome $\BMI \le \ry$ vs.~$\BMI > \ry$. The
transformation function $\h(\ry \mid \sex)$ is now a sex-specific intercept,
and $\shiftparm(\sex:\smoking)$ are the sex-specific log-odds ratios for the event $\BMI \le
\ry$ compared to the baseline category (never smokers). Because this shift
term does not depend on $\ry$, the model assumes proportionality of the
smoking odds with respect to the cut-off $\ry$. Stratification by sex allows
non-proportional smoking odds with respect to sex. In fact, the sex-specific conditional
distributions of males and females can still differ in very general ways
because two separate Bernstein polynomials $\h(\ry \mid \sex) =
\bern{5}^\top \parm(\sex)$ describe the conditional distributions for males and
females. The model can be seen as a
stratified proportional odds model for continuous responses or a continuous form
of logistic regression analyses, jointly performed for all possible cut-off
points $\ry$ under the assumption of constant  parameters
$\shiftparm(\sex:\smoking)$. Similar models, however without stratification,
were studied by \cite{Manuguerra_Heller_2010} using parametric intercept
functions $\h$ and recently by \cite{Liu_Shepherd_Li_2017} treating the
intercept function as a nuisance parameter in non-parametric maximum
likelihood estimation. \cite{Lohse_Rohrmann_Faeh_2017} provide a comparison
of parameter estimation in the presence of interval-censored body mass index
observations.

The parameterisation $\h(\ry \mid \sex) - \shiftparm(\sex:\smoking)$ with a
negative shift term seems unconventional from a logistic regression point of
view, but it simplifies interpretation. 
With model (\ref{mod:ss5}), $\E(\h(\rY \mid \sex, \smoking)) =
\shiftparm(\sex:\smoking)$, and thus positive shift parameters $\shiftparm$
indicate a shift of the BMI distribution towards higher BMI values.
Corresponding odds ratios $\exp(\shiftparm)$ larger than one mean that BMI
distributions are shifted to the right, compared to the BMI distribution in the
reference category.

Unfortunately, there was some further reduction in the log-likelihood
($-43639.74$), and interpretability doesn't come for free. However,
the model-based and empirical conditional BMI distribution functions look very much
the same as presented in Figure~\ref{ss4-plot} (additional plot not shown). The
sex-specific BMI-independent odds-ratios of smoking, compared to never
smoking, are given in Table~\ref{ss5-tab}. Former smokers had, on average,
a larger BMI compared to never smokers, and the effect was stronger for males. 
A similar effect was observed for male heavy smokers. Female light 
smokers showed a BMI distribution shifted to the left, compared with female
never smokers.

\begin{table}
\begin{center}
\caption{Linear transformation model (\ref{mod:ss5}) with sex and smoking-specific shift, stratified by sex. 
         Odds-ratios to the baseline category never
         smoking along with $95\%$ confidence intervals for males and females are shown.
         Odds ratios larger than one indicate a shift of the BMI distribution to the
         right. \label{ss5-tab}}
\begin{tabular}{llrr}
\toprule
 && \multicolumn{2}{c}{Sex}\\
\cmidrule{3-3}\cmidrule{4-4}
Smoking && \multicolumn{1}{c}{Female}&\multicolumn{1}{c}{Male}\\
\midrule
Never  &&                 1 &                 1\\
Former && 1.19 (1.08--1.31) & 1.95 (1.77--2.14)\\
Light  && 0.75 (0.65--0.85) & 0.95 (0.82--1.09)\\
Medium && 0.98 (0.85--1.12) & 0.93 (0.81--1.06)\\
Heavy  && 1.10 (0.92--1.32) & 1.43 (1.25--1.63)\\
\bottomrule
\end{tabular}

\end{center}
\end{table}

Maintaining interpretability, one could go further and assume equal smoking
effects for males and females in the model
\begin{eqnarray*}
\Prob(\BMI \le \ry \mid \sex, \smoking) = \expit\left(\h(\ry \mid \sex) -
\shiftparm(\smoking)\right).
\end{eqnarray*}

The log-likelihood was again reduced ($-43669.50$) for this model
with $16$ parameters. In addition, the
odds-ratios presented in Table~\ref{ss5-tab} indicate severe differences
in the smoking effects between males and females; therefore, I refrain from looking
at this or even simpler models and stop the top-down transformation choice
here. Of course, this very simple example only worked because it was possible to
compare models and raw data directly on the scale of the conditional BMI distribution
functions for two categorical explanatory variables, sex and smoking.
In the second part, I will consider additional, and also numeric,  
explanatory variables in a more realistic setup.

\section{Conditional BMI Distributions} \label{section:ctm}

My aim is to estimate the conditional BMI distribution given sex, smoking,
age and the lifestyle variables alcohol intake, education, physical
activity, fruit and vegetables consumption, residence and nationality as
explanatory variables $\rx$.  In the conditional transformation model
\begin{eqnarray*}
\Prob(\BMI \le \ry \mid \sex, \smoking, \age, \rx) = \expit\left(\h(\ry \mid \sex, \smoking, \age, \rx)\right),
\end{eqnarray*}
the conditional transformation function $\h$ depends on these variables in a
yet unspecified way. Top-down transformation choice ideally allows one to 
start without too many headaches, \ie an algorithm for fitting this model to handle the potentially many
explanatory variables of mixed type allows relatively complex
non-linear transformation functions. Such a model can be written as
\begin{eqnarray*}
\Prob(\BMI \le \ry \mid \sex, \smoking, \age, \rx) = \expit\left(\bern{5}(\ry)^\top \parm(\sex, \smoking, \age, \rx)\right),
\end{eqnarray*}
assuming that each conditional distribution is parameterised in terms of a
Bernstein polynomial of order $5$. The parameters $\parm$ of
this polynomial, however, depend on the explanatory variables in a
potentially complex way, featuring interactions and non-linearities. 
Tree and forest algorithms \citep{Hothorn_Zeileis_2017} allow such
``conditional parameter functions'' $\parm(\sex, \smoking, \age,
\rx)$, and thus the corresponding conditional BMI distributions, to be
estimated in a black-box manner without the necessity to \apriori specify 
any structure of $\parm(\sex, \smoking, \age, \rx)$. I will first use
trees and forests to understand the complexity of the impact of the
explanatory variables on the BMI distribution. Later on, I will apply a
top-down approach to transformation choice to obtain simpler transformation
models that allow more straightforward interpretation.

\subsection{Transformation Trees and Forests}

\begin{sidewaysfigure}
\begin{knitrout}
\definecolor{shadecolor}{rgb}{0.969, 0.969, 0.969}\color{fgcolor}
\includegraphics[width=\maxwidth]{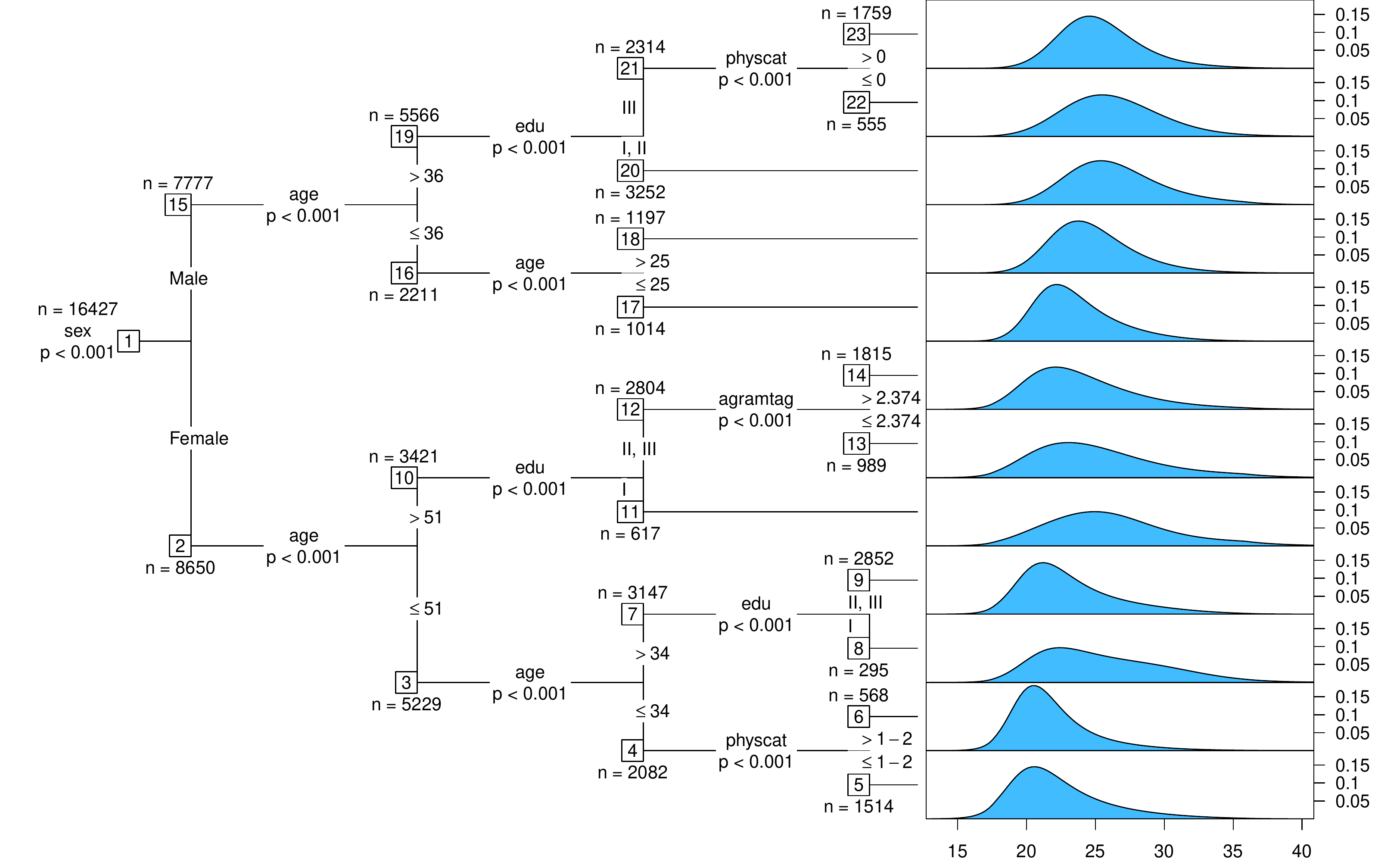} 

\end{knitrout}
\caption{Transformation tree. The conditional BMI distributions (depicted in
         terms of their densities) are given in each subgroup corresponding
         to the terminal nodes of the tree. Variables: education (edu) at
         levels mandatory (I), secondary (II) and tertiary (III); 
         alcohol intake (agramtag). 
         \label{cmpx-tree-plot}}
\end{sidewaysfigure}

A transformation tree \citep{Hothorn_Zeileis_2017} starts with an unconditional
transformation model
\begin{eqnarray} \label{mod:unc}
\Prob(\BMI \le \ry) = \expit\left(\bern{5}(\ry)^\top \parm\right)
\end{eqnarray}
and a corresponding maximum-likelihood estimator
$\hat{\parm}$. The algorithm proceeds by assessing correlations between 
the score contributions evaluated at 
$\hat{\parm}$ and the explanatory variables sex, smoking, age and $\rx$. A binary split is
implemented in the most discriminating cut-off point of the variable showing
the highest correlation to any score.  The procedure is repeated until a
certain stop criterion applies.  The result is a partition of the data. The
algorithm is sensitive to distributional changes, \ie the
conditional BMI distributions in the subgroups of this partition may vary
with respect to the mean BMI and also with respect to higher BMI moments. 
In each subgroup, the unconditional model~(\ref{mod:unc}) was used to estimate
$\parm(\sex, \smoking, \age, \rx)$ for this subgroup.  Because each
observation in this subgroup is then associated with a dedi\-cated parameter
vector $\hat{\parm}(\sex, \smoking, \age, \rx)$, the
log-likelihood for the tree model could be evaluated as the sum of the likelihoods in the
subgroups.  The log-likelihood of the tree presented in
Figure~\ref{cmpx-tree-plot} is $-43079.42$.  The first split
is in sex, so in fact two sex-specific models are given here.  Three age
groups ($\le 34$, $(34, 51]$, $> 51$) for females and three age
groups ($\le 25$, $(25, 36]$, $> 36$) for males are distinguished. 
Education contributed to understanding the BMI distribution of females and
males.  Location, scale and
shape of the conditional BMI distributions varied considerably.  The variance
increased with age, and higher-educated people tended to have lower BMI
values.  These are interesting insights, but the model is of course very
rough.

\begin{figure}[t]
\begin{knitrout}
\definecolor{shadecolor}{rgb}{0.969, 0.969, 0.969}\color{fgcolor}
\includegraphics[width=\maxwidth]{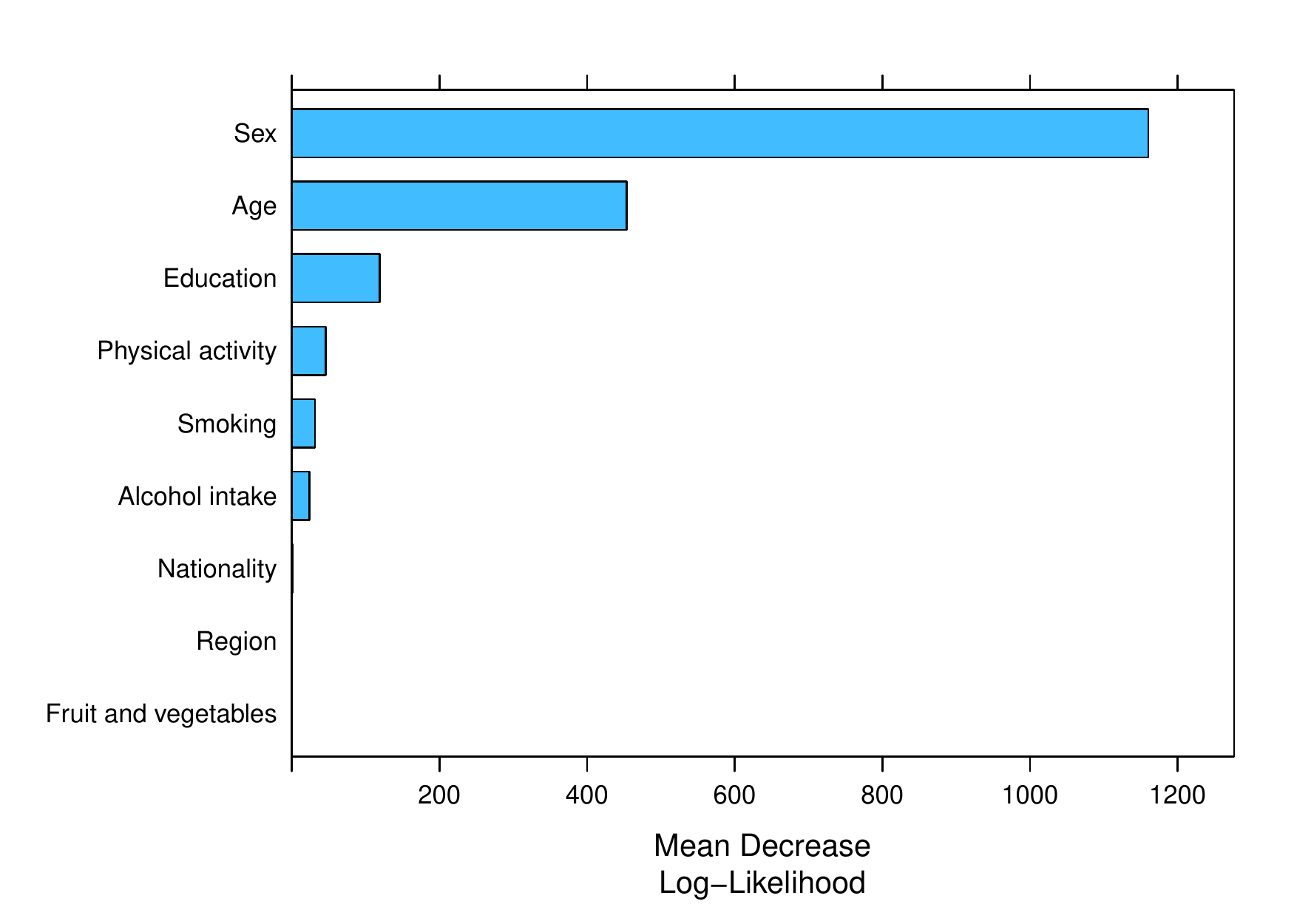} 

\end{knitrout}
\caption{Transformation forest. Likelihood-based permutation variable importance
         for all variables in the forest. The $x$-axis shows the mean
         decrease in the log-likelihood caused by permuting one variable
         before computing the in-sample tree log-likelihood. \label{cmpx-varimp-plot}}
\end{figure}


A transformation forest \citep{Hothorn_Zeileis_2017} allows less rough
conditional parameter functions $\parm(\sex, \smoking, \age, \rx)$ to be
estimated. There are no longer any restrictions regarding the conditional parameter
functions. In this sense, a transformation forest is the ``most complex model
one can think of'' as mentioned in the introduction. 
The random forest class of models is considered  to be very accurate, insensitive to
hyperparameter tuning and without a tendency to overfit. In the following, I
shall use this method to obtain a benchmark for better-interpretable
transformation models following the top-down model selection approach.

The generic random forest algorithm essentially relies on 
multiple transformation trees fitted to subsamples of the data, with a random
selection of variables to be considered for splitting in each node.
Unlike the original random forest \citep{Breiman_2001}, a transformation
model can be understood as a procedure assigning a parametric model to each
observation. For subject $i$, the forest conditional distribution
function is
\begin{eqnarray*}
\hat{\Prob}(\BMI \le \ry \mid \sex_i, \smoking_i, \age_i, \rx_i) = \expit\left(\bern{5}(\ry)^\top
\hat{\parm}(\sex_i, \smoking_i, \age_i, \rx_i)\right).
\end{eqnarray*}
In this sense, a
transformation forest ``predicts'' a fully parametric model for each
subject, albeit with a very flexible conditional parameter.
The conditional parameter $\hat{\parm}(\sex_i, \smoking_i, \age_i, \rx_i)$ was obtained
from a locally adaptive maximum-likelihood estimator based on so-called
nearest neighbour weights \citep{Hothorn_Zeileis_2017}. 
A considerable improvement in the transformation forest log-likelihood
($-42520.18$) was observed.  In fact, this is the largest log-likelihood
I was able to achieve. Thus, this transformation forest is the best-
fitting model for the BMI data. 

\begin{figure}[t]
\begin{knitrout}
\definecolor{shadecolor}{rgb}{0.969, 0.969, 0.969}\color{fgcolor}
\includegraphics[width=\maxwidth]{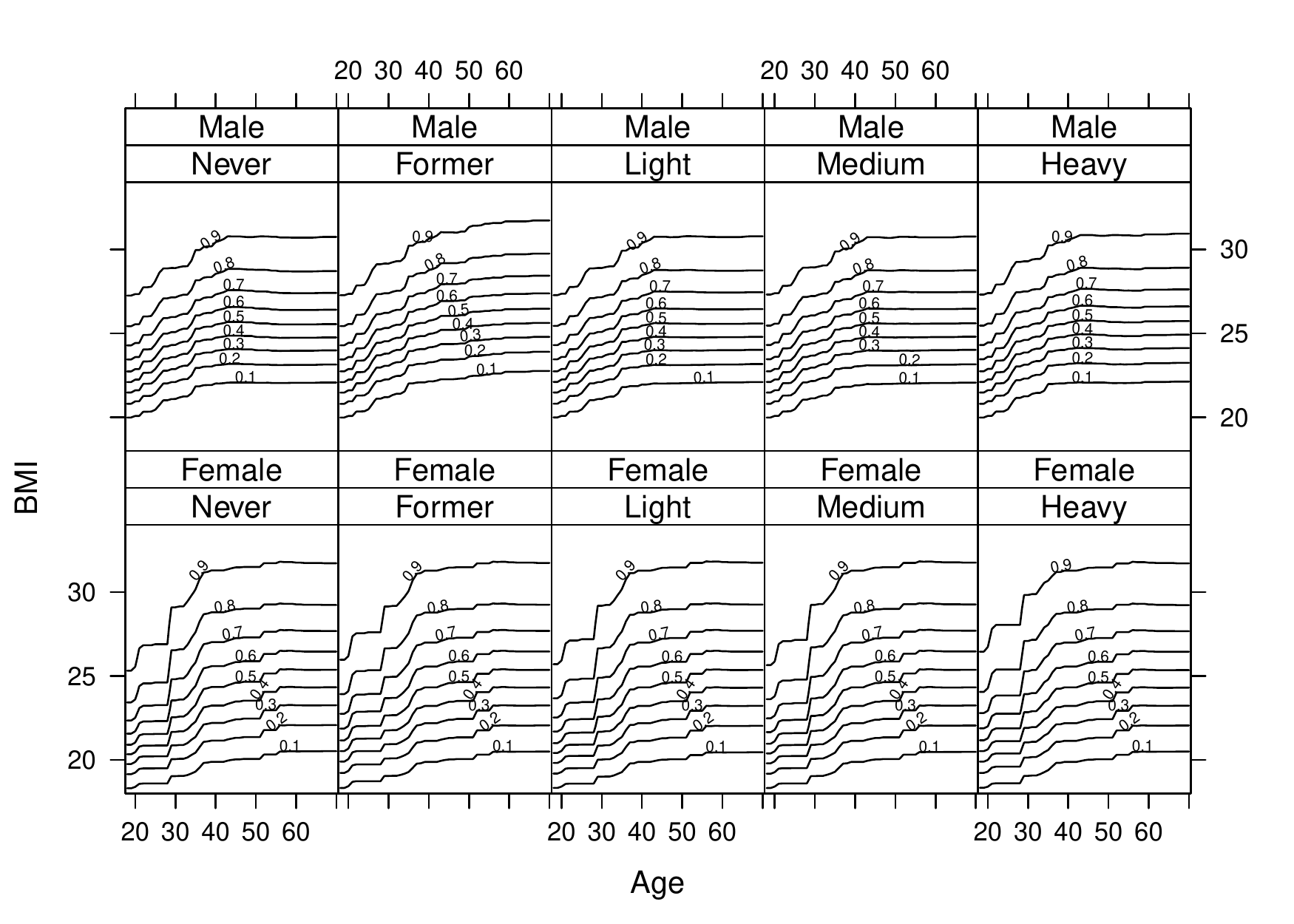} 

\end{knitrout}
\caption{Transformation forest. Partial dependency of sex, smoking and age.
         Conditional decile curves for BMI depending on age, separately
         for all combinations of sex and smoking, are shown.
         \label{cmpx-age-plot}}
\end{figure}

On the downside, this black-box model makes is very
difficult to understand the impact of the explanatory variables on the
conditional BMI distribution.  The likelihood-based permutation variable importance
(Figure~\ref{cmpx-varimp-plot}) indicated that only sex, age, education,
physical activity and smoking have an impact on BMI, where again sex seems to be the
most important variable.  Age was a more important factor than education or
physical activity, and thus the only numeric variable one needs to consider.
The association between sex, smoking, age and BMI as
described by the transformation forest is given in terms of a partial
dependency plot of conditional deciles in Figure~\ref{cmpx-age-plot}.  In
general, the median BMI increases with age, as does the BMI variance.  For males,
there seemed to be a level-effect whose onset depends on smoking category. 
Females tended to higher BMI values, and the variance was larger compared to
males.  There seemed to be a bump in BMI values for 
females, roughly around $30$ years. This
corresponds to mothers giving birth to their first child around this age. 
It is important to note that the right-skewness of the
conditional BMI distributions in Figure~\ref{cmpx-age-plot} renders 
conditional normal distributions inappropriate, even under variance
heterogeneity.

This complex model would be sufficient if one was only interested in the
estimation of conditional BMI distributions for persons with specific
configurations of the sex, smoking, age and the remaining explanatory
variables. The variable importances can be used to rank variables according
to their impact on the conditional BMI distributions but cannot replace
effect measures, let alone an assessment of their variability. Communication
with subject-matter scientists and publication of results in subject-matter
journals requires simplification of these models. Top-down transformation
choice can help to find models of appropriate complexity, as will be seen in the
next section.

\subsection{Conditional Transformation Model}

The analysis using transformation trees and especially transformation
forests revealed strong effects of sex and age; the latter variable
was not considered in our analysis presented in
Section~\ref{section:sexsmoking}. A more structured model roughly as 
powerful as the transformation forest must therefore allow the
conditional distribution of BMI to change with both sex and age in 
very general ways. The remaining variables were less important, and one can
hopefully cut some corners here by assuming simple linear main effects for
these variables. I start the top-down search for a simpler model with
a conditional transformation model of the form
\begin{eqnarray} \label{mod:ctm}
& & \Prob(\BMI \le \ry \mid \sex, \smoking, \age, \rx) = \\ \nonumber
& & \quad \expit\left(\h(\ry, \age \mid \sex) -  \shiftparm(\sex:\smoking) - \rx^\top \shiftparm \right).
\end{eqnarray}
The transformation function $\h(\ry, \age \mid \sex)$ implements a
sex-specific bivariate smooth-surface function of BMI and $\age$, which was of course 
monotonic in its first argument. The surface function for males explains
age-induced changes in the conditional distribution of BMI. In contrast to
transformation forests, the assumption was made that the function is smooth in both $\ry$
and $\age$ and not only in $\ry$.  I parameterise this function as
a tensor product of two Bernstein polynomials of order 5, one
for BMI and one for age, with sex-specific $36$-dimensional parameter vector $\parm(\sex)$, in other words as 
$(\bern{5}(\ry) \otimes \bern{5}(\age))^\top
\parm(\sex)$. Except for $\smoking$, the remaining variables
entered only as the linear shift term $\rx^\top \shiftparm$ of main effects.
In light of its fifth rank in the permutation variable importance
(Figure~\ref{cmpx-varimp-plot}), it may seem a bit inconsequent to treat 
$\smoking$ differently than the other variables. However, the stratified
analysis in Section~\ref{section:sexsmoking} suggested the need for
sex-specific smoking effects, and I thus include the interaction term
$ \shiftparm(\sex:\smoking)$ also in this model. The $\expit$ function
around the transformation function ensures interpretability of 
all regression coefficients $\shiftparm$ on the log-odds scale.

\begin{figure}[t]
\begin{knitrout}
\definecolor{shadecolor}{rgb}{0.969, 0.969, 0.969}\color{fgcolor}
\includegraphics[width=\maxwidth]{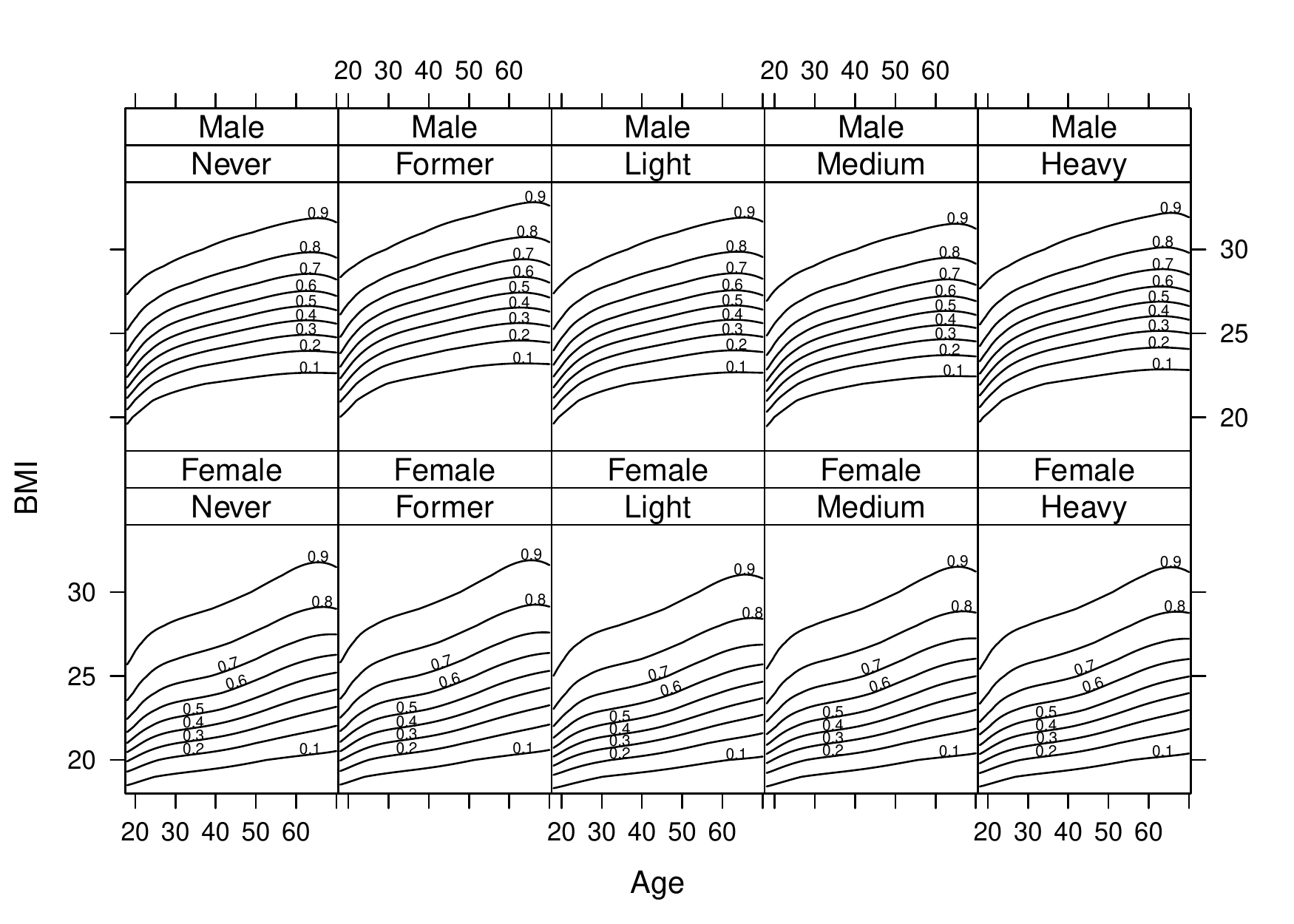} 

\end{knitrout}
\caption{Conditional transformation model (\ref{mod:ctm}).
         Conditional decile curves for BMI depending on age, separately
         for all combinations of sex and smoking, are shown. \label{cmpx-ctm-plot}}
\end{figure}

With $89$ parameters, the log-likelihood
$-42778.14$ of model (\ref{mod:ctm}) was only slightly smaller
than the log-likelihood of the transformation forest
($-42520.18$).  In a certain sense, this conditional
transformation model can be seen as an approximation of the black-box
transformation forest.  The effects of sex, smoking and age, with all
remaining variables being constant, are again best visualised using the
conditional decile functions (Figure~\ref{cmpx-ctm-plot}).  The decile
functions are now smooth in age due to the parameterisation of the age
effect in terms of Bernstein polynomials.  For males, the BMI increased with
age; the BMI reduction in males older than $65$ years was not visible in the
decile curves of the transformation forests (Figure~\ref{cmpx-age-plot}). 
The slope was largest for young men up to $25$ years, followed by a linear
increase until the age of $65$.  The male BMI distribution was right skewed,
with only a small increase in the variance towards older people.  For
females, a bump in the BMI distribution was again identified around the
age of $30$, corresponding to pregnancies and breast-feeding times. 
The effect seemed more pronounced in higher deciles.  Right skewness and a
variance increase towards older women can be inferred from this figure.

The main advantage of this complexity reduction is the interpretability of
the regression coefficients $\shiftparm(\sex:\smoking)$ and $\shiftparm$
in terms of BMI-independent log-odds ratios. The sex-specific smoking
effects and the effects of the remaining variables as odds ratios are given
in the left column of Table~\ref{summary-tab}. Further simplification can be
achieved by replacing the bivariate surface function of $\ry$ and age by
a sex-specific, BMI-varying linear effect of age in the distribution
regression model presented in the next section.

\subsection{Distribution Regression}

The term ``distribution regression'' \citep{Chernozhukov_2013} is commonly 
used to describe response-varying
coefficients. In survival analysis, the term ``time-varying coefficients'' is
more typical. Here, a BMI-varying coefficient of age is a means of simplifying the
conditional transformation model (\ref{mod:ctm}). In the simpler model, I assume a smoothly varying but sex-specific 
coefficient of age $\eshiftparm(\ry | \sex)$. The transformation function
$\h(\ry \mid \sex)$ is again the simple transformation of BMI given sex
introduced in model (\ref{mod:ss4}). The model reads
\begin{eqnarray} \label{mod:dr}
& & \Prob(\BMI \le \ry \mid \sex, \smoking, \age, \rx) = \\ \nonumber
& & \quad \expit\left(\h(\ry \mid \sex) - \eshiftparm(\ry | \sex )\age - 
        \shiftparm(\sex:\smoking) - \rx^\top \shiftparm\right).
\end{eqnarray}

\begin{figure}[t]
\begin{knitrout}
\definecolor{shadecolor}{rgb}{0.969, 0.969, 0.969}\color{fgcolor}
\includegraphics[width=\maxwidth]{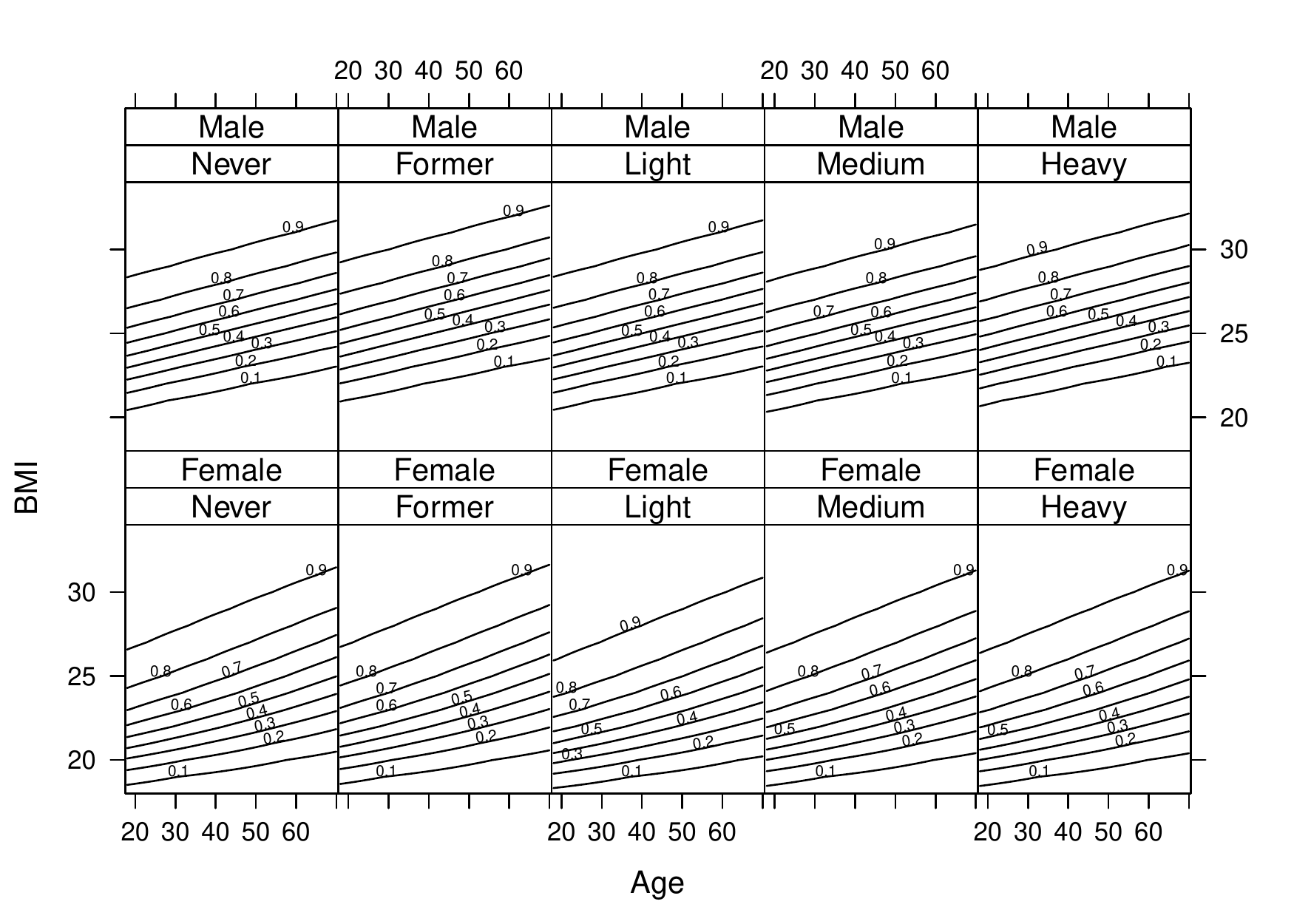} 

\end{knitrout}
\caption{Distribution regression model (\ref{mod:dr}).
         Conditional decile curves for BMI depending on age, separately
         for all combinations of sex and smoking, are shown. \label{dr-plot}}
\end{figure}

The log-likelihood $-42888.10$ decreased considerably in this model
with $41$ parameters. The effects of smoking and
the remaining variables (except age) are given in the middle column of
Table~\ref{summary-tab} as odds ratios. When the dependency of BMI
deciles on sex, smoking and age were depicted (Figure~\ref{dr-plot}), the linear structure
regarding age was obvious. The age-varying slopes and the pregnancy bump could
not be identified by this simpler model. Right-skewness and variance
heterogeneity for females remained visible. The variance increase in older
males now seemed questionable. For my taste, the replacement of two
bivariate functions by two univariate functions does not really help model
interpretation, as one would have to plot these two functions in any case. The
severe reduction of the log-likelihood indicated that the effect of age is better
described in a conditional transformation model of the form (\ref{mod:ctm}).
Nevertheless, I will go one step further and connect the stratified 
linear transformation model (\ref{mod:ss5}) with a model of the same form 
featuring age and the lifestyle variables $\rx$ in addition to sex and
smoking.

\subsection{Stratified Linear Transformation Model}

I extend the stratified linear transformation model (\ref{mod:ss5}) with an
sex-specific age effect and a linear predictor $\rx^\top \shiftparm$ of the
remaining variables
\begin{eqnarray} \label{mod:smpl}
& & \Prob(\BMI \le \ry \mid \sex, \smoking, \age, \rx) = \\ \nonumber
& & \quad \expit\left(\h(\ry \mid \sex) - 
    \shiftparm(\sex:\smoking) - \shiftparm(\sex:\age) - \rx^\top \shiftparm\right)
\end{eqnarray}

The log-likelihood was further reduced to $-42900.36$.  In this
model, the sex differences in the age effects were completely gone, as the
odds ratios for a one-year increase were $1.03$ ($1.03-1.04$) for males and
$1.03$ ($1.03-1.04$) for females.  In light of the more complex structure of
the age effect identified by the more complex models, one would incorrectly
draw the conclusion of equal age effects for males and females based on this
oversimplified model.  The effects of the remaining parameters are given in
the right column of Table~\ref{summary-tab}.

\begin{table}[t]
\caption{Transformation models (\ref{mod:ctm}--\ref{mod:smpl}). 
         Odds ratios with $95\%$ confidence intervals. 
         Values larger than one indicate a shift of the BMI distribution to the
         right (compared to the baseline category or a one-unit increase in
         a numeric variable). \label{summary-tab}}
\begin{tabular}{lrrr}
\toprule
 & \multicolumn{1}{c}{Model (\ref{mod:ctm})} & \multicolumn{1}{c}{Model (\ref{mod:dr})} & \multicolumn{1}{c}{Model (\ref{mod:smpl})} \\
\midrule
Smoking (Females) &                   &                   &                   \\
\hspace{.4cm}Never &               $1$ &               $1$ &               $1$ \\
\hspace{.4cm}Former & 1.04 (0.95--1.15) & 1.06 (0.96--1.16) & 1.06 (0.96--1.16) \\
\hspace{.4cm}Light & 0.81 (0.70--0.92) & 0.81 (0.71--0.93) & 0.81 (0.71--0.93) \\
\hspace{.4cm}Medium & 0.92 (0.80--1.06) & 0.94 (0.82--1.08) & 0.94 (0.82--1.08) \\
\hspace{.4cm}Heavy & 0.91 (0.76--1.09) & 0.93 (0.78--1.12) & 0.94 (0.78--1.12) \\
Smoking (Males) &                   &                   &                   \\
\hspace{.4cm}Never &               $1$ &               $1$ &               $1$ \\
\hspace{.4cm}Former & 1.44 (1.31--1.59) & 1.47 (1.33--1.62) & 1.47 (1.33--1.61) \\
\hspace{.4cm}Light & 1.02 (0.88--1.17) & 1.01 (0.88--1.16) & 1.01 (0.88--1.16) \\
\hspace{.4cm}Medium & 0.87 (0.76--1.00) & 0.90 (0.79--1.03) & 0.91 (0.79--1.04) \\
\hspace{.4cm}Heavy & 1.13 (0.99--1.30) & 1.21 (1.06--1.39) & 1.22 (1.07--1.40) \\
Alcohol intake (g/d) & 1.00 (1.00--1.00) & 1.00 (1.00--1.00) & 1.00 (1.00--1.00) \\
Fruit and vegetables &                   &                   &                   \\
\hspace{.4cm}High &               $1$ &               $1$ &               $1$ \\
\hspace{.4cm}Low & 1.07 (1.01--1.13) & 1.08 (1.02--1.14) & 1.08 (1.02--1.14) \\
Physical activity &                   &                   &                   \\
\hspace{.4cm}High &               $1$ &               $1$ &               $1$ \\
\hspace{.4cm}Moderate & 1.11 (1.04--1.19) & 1.15 (1.08--1.23) & 1.16 (1.08--1.23) \\
\hspace{.4cm}Low & 1.25 (1.16--1.34) & 1.30 (1.21--1.40) & 1.30 (1.21--1.40) \\
Education &                   &                   &                   \\
\hspace{.4cm}Mandatory (I) &               $1$ &               $1$ &               $1$ \\
\hspace{.4cm}Secondary (II) & 0.72 (0.66--0.79) & 0.79 (0.73--0.87) & 0.80 (0.73--0.87) \\
\hspace{.4cm}Tertiary (III) & 0.48 (0.43--0.52) & 0.56 (0.51--0.61) & 0.56 (0.51--0.62) \\
Nationality &                   &                   &                   \\
\hspace{.4cm}Swiss &               $1$ &               $1$ &               $1$ \\
\hspace{.4cm}Foreign & 1.17 (1.09--1.25) & 1.23 (1.15--1.31) & 1.24 (1.16--1.32) \\
Region &                   &                   &                   \\
\hspace{.4cm}German speaking &               $1$ &               $1$ &               $1$ \\
\hspace{.4cm}French speaking & 0.89 (0.83--0.94) & 0.88 (0.83--0.94) & 0.88 (0.83--0.94) \\
\hspace{.4cm}Italian speaking & 0.81 (0.72--0.93) & 0.81 (0.71--0.92) & 0.81 (0.71--0.92) \\
\bottomrule
\end{tabular}

\end{table}

The three columns presented in Table~\ref{summary-tab} refer to the same
parameters, estimated by three models differing only with respect to the
complexity of the age effect.  The effects of smoking, alcohol intake,
education, physical activity, fruit and vegetables consumption, residence
and nationality were remarkably constant.  Alcohol intake had no impact on
the BMI in this study, and right shifts in BMI distributions were associated
with low fruit and vegetable consumption, moderate and low physical
activity, short education, being a foreigner or living in the 
German-speaking part of Switzerland.  These conclusions can be drawn from all three
models in the same way.  The effects of smoking were less pronounced than
the effects obtained in our initial analysis that ignored age and the
lifestyle variables (Table~\ref{ss5-tab}).  Light smokers had lower BMIs
than never smokers; the remaining effects are questionable.

\section{Discussion}

The core of top-down transformation choice is a family of decreasingly
complex, yet fully comparable, conditional transformation models.  Model parameterisation
and interpretation in the family of transformation models are always 
based on the conditional distribution function
\begin{eqnarray*}
\pBMI(\ry \mid \sex, \smoking, \age, \rx) = F(\h(\ry \le \ry \mid \sex, \smoking, \age, \rx)).
\end{eqnarray*}
Unlike most classical models featuring explicit parameters for conditional
means or conditional variances, transformation models describe conditional
distributions explicitly and moments implicitly. What might seem as a
disadvantage is in fact, as I hope I could convince the readers of, a very
attractive feature of transformation models for regression analysis. In this
tutorial, I exclusively defined and interpreted models for conditional
distributions. The corresponding distribution functions were used to 
compare transformation models with the
empirical cumulative distribution function (Figures~\ref{ss1-plot},
\ref{ss2-plot}, \ref{ss3-plot} and \ref{ss4-plot}). The conditional
transformation function $\h$
was used to assess deviations from normality in Figure~\ref{ss3-trafo-plot}.
Conditional densities
\begin{eqnarray*}
\dBMI(\ry \mid \sex, \smoking, \age, \rx) = \pBMI^\prime(\BMI \le \ry \mid \sex, \smoking, \age, \rx)
\end{eqnarray*}
are depicted in Figure~\ref{ss3-plot-density} and those for each terminal node of the transformation tree are shown in
Figure~\ref{cmpx-tree-plot}. Densities defined the log-likelihood
\begin{eqnarray*}
\sum_{i = 1}^{16{,}427} w_i \log(\dBMI(\ry_i \mid \sex_i, \smoking_i, \age_i, \rx_i))
\end{eqnarray*}
based on all $16{,}427$ BMI measurements $\ry_i$ with sampling weights
$w_i$. Conditional quantile functions
\begin{eqnarray*}
\qBMI(p \mid \sex, \smoking, \age, \rx) = \pBMI^{-1}(p \mid \sex, \smoking, \age, \rx)
\end{eqnarray*}
helped to visualise age effects in Figures~\ref{cmpx-age-plot},
\ref{cmpx-ctm-plot} and \ref{dr-plot}. Effect measures for sex, smoking
and lifestyle variables in Tables~\ref{ss5-tab} and \ref{summary-tab}
were obtained as ratios of conditional odds functions
\begin{eqnarray*}
\oBMI(\ry \mid \sex, \smoking, \age, \rx) = \frac{\pBMI(\ry \mid \sex, \smoking, \age, \rx)}{1 - \pBMI(\ry \mid \sex, \smoking, \age,
\rx)}.
\end{eqnarray*}
Varying model complexity only affects the flexibility of these functions that
characterise conditional distributions, but not the corresponding
interpretations.

A unique feature of conditional transformation models is the ability
to formulate, estimate, compare, evaluate, interpret and understand models
seemingly as far apart as a normal linear model with constant variance and a
transformation forest in the same theoretical framework.  Straightforward
answers to some questions that have plagued data analysis for decades, for
example ``Is it appropriate to assume normal errors?'' or ``How should the
response be transformed prior to analysis?'',
are easily obtained from conditional transformation models.

One practical and interesting question relates to the impact of the order $M$
of the Bernstein polynomial $\bern{M}(\ry)^\top \parm$.  The choice $M = 1$
implements a linear function, and with $F = \Phi$, conditional
normal distributions are obtained. For $M \rightarrow \infty$,
$\bern{M}(\ry)^\top \parm$ converges uniformly to the true and unknown
transformation function $\h(\ry)$ in a model $\Prob(\rY \le \ry) =
F(\h(\ry))$.  Because $\h$ is a monotonic function, too-erratic behaviour
cannot occur, even for very large $M$, and overfitting is not an issue
\citep[see][for numerical examples]{vign:mlt.docreg}.  In the
model~(\ref{mod:ss3}), increasing the order from $M = 5$ to $M =
10$ led to a very small increase in the log-likelihood
from $-43564.30$ to $-43547.16$. In the extreme case of
very large $M$, the conditional distribution function $F(\bern{M}(\ry)^\top \parm)$
closely interpolates the empirical cumulative distribution function. The
latter estimator is consistent, as is the transformation model
\citep{Hothorn_Moest_Buehlmann_2016}.

This tutorial did not address any issue regarding model estimation or model
inference.  Details about maximum-likelihood estimation in conditional
transformation models can be found in \cite{Hothorn_Moest_Buehlmann_2016}. 
Locally adaptive maximum-likelihood estimation for transformation trees and
transformation forests has been introduced in \cite{Hothorn_Zeileis_2017}. More
elaborate discussions of model parameterisation in conditional
transformation models and of connections to other models can be found in 
\cite{Hothorn_Kneib_Buehlmann_2014} and \cite{Hothorn_Moest_Buehlmann_2016}.
Applications of conditional transformation models can be found in 
\cite{Hothorn_Kneib_Buehlmann_2013}, \cite{Moest_Schmid_Faschingbauer_2014}
and \cite{Moest_Hothorn_2015}. An introduction to the \pkg{mlt} 
add-on package \citep{pkg:mlt} for maximum-likelihood estimation in
conditional transformation models, including models for ordinal or censored
and truncated responses, is available in \cite{vign:mlt.docreg}.

\section*{Reproducibility}

Data from the Swiss Health Survey 2012 can be obtained from the Swiss
Federal Statistics Office (Email: \url{sgb12@bfs.admin.ch}).  Data is available
for scientific research projects, and a data protection application form
must be submitted.  More information can be found here
\url{http://www.bfs.admin.ch/bfs/de/home/statistiken/gesundheit/erhebungenSupplementary}.

The code used for producing the results presented in this paper can be
evaluated on a smaller artificial data set sampled from the transformation
forest by running \texttt{demo("BMI")} from the \pkg{trtf} package
\citep{pkg:trtf}.

\section*{Acknowledgements}

I thank the students participating in the course ``STA660 Advanced R
Programming'' that I taught in the spring semester of 2017 for producing the code
underlying Figure~\ref{cmpx-tree-plot} as part of their homework assignments. 
Parts of this paper were written during a research sabbatical at Universit\"at Innsbruck
financially supported by the Swiss National Science Foundation (grant number
IZSEZ0\_177091).

\clearpage

\bibliography{mlt,add,packages}

\end{document}